\documentclass[review]{elsarticle}

\let\today\relax
\makeatletter
\def\ps@pprintTitle{%
    \let\@oddhead\@empty
    \let\@evenhead\@empty
    \def\@oddfoot{\footnotesize\itshape
         {Submitted preprint} \hfill\today}%
    \let\@evenfoot\@oddfoot
    }
\makeatother

\usepackage{lineno,hyperref}
\usepackage{graphics}
\usepackage{hyperref}

\biboptions{authoryear}

\begin{document}

\begin{frontmatter}

\title{WHOCARES: data-driven WHOle-brain CArdiac signal REgression from highly accelerated simultaneous multi-Slice fMRI acquisitions}

\author[CE]{Nigel Colenbier\fnref{myfootnote1,myfootnote2,myfootnote3}}
\author[CE]{Marco Marino\fnref{myfootnote1,myfootnote2}}
\author{Giorgio Arcara\fnref{myfootnote1}}
\author{Blaise Frederick\fnref{myfootnote4,myfootnote5}}
\author{Giovanni Pellegrino\fnref{myfootnote1}}
\author[LA]{Daniele Marinazzo\fnref{myfootnote1,myfootnote3}}
\author[LA,san_camillo]{Giulio Ferrazzi\corref{mycorrespondingauthor}\fnref{myfootnote1}}
\fntext[myfootnote1]{IRCCS San Camillo Hospital, via Alberoni 70, 30126 Venice, Italy}
\fntext[myfootnote2]{Research Center for Motor Control and Neuroplasticity, KU Leuven, Leuven, 3001, Belgium}
\fntext[myfootnote3]{Department of Data Analysis, Faculty of Psychology and Educational Sciences, Ghent University, Ghent 9000, Belgium }
\fntext[myfootnote4]{Brain Imaging Center, McLean Hospital, 115 Mill St., Belmont, MA, 02478, USA}
\fntext[myfootnote5]{Department of Psychiatry, Harvard University Medical School, 25 Shattuck St., Boston, MA, 02115, USA}

\cortext[mycorrespondingauthor]{Corresponding author}
\ead{giulio.ferrazzi@hsancamillo.it}
\address[CE]{Contributed equally to the paper}
\address[LA]{Joint last authors}
\address[san_camillo]{via Alberoni 70, 30126 Venice, Italy}

\begin{abstract}

Cardiac pulsation is a physiological confound of functional magnetic resonance imaging (fMRI) time-series that introduces spurious signal fluctuations in proximity to blood vessels. fMRI alone is not sufficiently fast to resolve cardiac pulsation. Depending on the ratio between the instantaneous heart-rate and the acquisition sampling frequency ($\frac{1}{TR}$, with $TR$ being the repetition time), the cardiac signal may alias into the frequency band of neural activation. The introduction of simultaneous multi-slice (SMS) imaging has significantly reduced the chances of cardiac aliasing. However, the necessity of covering the entire brain at high spatial resolution restrain the shortest $TR$ to just over 0.5 seconds, which is in turn not sufficiently short to resolve cardiac pulsation beyond 60 beats per minute. Recently, hyper-sampling of the fMRI time-series has been introduced to overcome this issue. While each anatomical location is sampled every $TR$ seconds, the time between consecutive excitations is shorter and thus adequate to resolve cardiac pulsation. In this study, we show that it is feasible to temporally and spatially resolve cardiac waveforms at each voxel location by combining a dedicated hyper-sampling decomposition scheme with SMS. We developed the technique on 774 healthy subjects selected from the Human Connectome Project (HCP) and validated the technology against the RETROICOR method. The proposed approach, which we name Data-driven WHOle-brain CArdiac signal REgression from highly accelerated simultaneous multi-Slice fMRI acquisitions (WHOCARES), is fully data-driven, does not make specific assumptions on cardiac pulsatility, and is independent from external physiological recordings so that the retrospective correction of fMRI data becomes possible when such measurements are not available. WHOCARES is freely available at \href{https://github.com/gferrazzi/WHOCARES}{https://github.com/gferrazzi/WHOCARES}.

\end{abstract}


\end{frontmatter}

\section*{Highlights} \label{section:Highlights}

\begin{itemize}
  \item A data-driven method to perform cardiac signal regression in fMRI is presented
  \item The technique is tested on 774 healthy subjects from the Human Connectome Project
  \item The technology is validated against the state-of-the-art RETROICOR method
  \item The approach is independent from external physiological recordings
  \item The approach does not impose modeling priors on the shape of the cardiac regressor
\end{itemize}

\paragraph{Non standard abbreviations}\footnote{cardiac part of RETROICOR (\textbf{Cardio-RETROICOR}), data-driven WHOle-brain CArdiac signal REgression from highly accelerated simultaneous multi-Slice fMRI acquisitions (\textbf{WHOCARES}), heart-rate (\textbf{HR}), heart-rate variability (\textbf{HRV}), median absolute deviation (\textbf{MAD}), number of temporal volumes (\textbf{NV}), quality control (\textbf{QC}), time between consecutive excitations (\textbf{TS}).}

\section{Introduction} \label{section:Introduction}

The field of functional neuroimaging has seen rapid growth since the discovery of the blood-oxygenation-level-dependent (BOLD) contrast (\cite{ogawa1990brain}). Functional magnetic resonance imaging (fMRI) is the workhorse modality for BOLD imaging due to its sensitivity to local magnetic field changes caused by the presence/absence of oxygenated/deoxygenated blood. However, fMRI is subject to artefacts deriving from multiple sources including motion (\cite{power2015recent}) and physiological sources (\cite{kruger2001physiological, triantafyllou2005comparison}). 

Systemic physiological changes, including those associated with cardiac and respiratory cycles, are well known to influence hemodynamic changes through multiple mechanisms. In particular, low-frequency fluctuations of arterial carbon dioxide (\cite{wise2004resting}), breathing (\cite{birn2006separating, raj2001respiratory, windischberger2002origin}), heart-rate (HR) and heart-rate variability (HRV) (\cite{chang2009influence, shmueli2007low}) account for considerable variance both in resting state as well as during tasks. Recent evidence suggests that physiological contributions cannot be removed by the use of a single nuisance regressor, such as global signal regression (\cite{erdougan2016correcting, liu2017global}), due to the regional heterogeneity of their responses (\cite{chen2020resting}).

An interesting feature of fMRI acquisitions is that physiological noise can be resolved locally if a sufficiently fast acquisition is employed. Appealing as it may seem, this approach is typically not used since the technological constrains of fMRI limit the highest possible sampling rate. In fact, the time between the acquisition of two consecutive slices (hereafter referred to as $TS$) varies between 50$ms$ to 100$ms$ depending on resolution, field of view, gradients performance, BOLD contrast (\cite{ferrazzi2016exploration}), safety and acceleration requirements. Since several slices (hereafter referred to as $Z$ slices) are needed to cover the brain with margins for motion, and given that slices are typically acquired in a sequential manner, the repetition time (i.e. the time it takes for the same anatomical location to be sampled successively in time, or $TR$) can last several seconds. If we assume a typical respiration pattern of 12 to 15 breaths per minute, the maximum allowable $TR$ would be within the interval 2 to 2.5 seconds to avoid respiratory induced aliasing. For the case of cardiac pulsation, the requirements are even more stringent since a heartbeat of 50 to 70 beats per minute (BPM) would result in a maximum TR of 0.43 to 0.6 seconds. Moreover, cardiac and respiratory noise are semi-periodic functions and as such they may include higher-order harmonics not unambiguously resolved even when ultra-fast acquisition are employed (\cite{chen2019analysis}). 

A number of techniques have been introduced for the acceleration of fMRI sequences. Perhaps, the most efficient method to achieve acceleration in fMRI is simultaneous multi-slice (SMS) imaging (\cite{larkman2001use}). In SMS, several slices are excited simultaneously employing a multiband ($MB$) radio-frequency (RF) pulse. Slices are shifted along the phase encoding direction with blipped controlled aliasing in parallel imaging results in higher acceleration (CAIPIRINHA) encoding (\cite{setsompop2012blipped}), and separated using the information enclosed within the sensitivity profiles of the multiple receiver coils (\cite{griswold2002generalized, pruessmann1999sense}). One of the most attractive features of SMS is that $TR$ is inversely proportional to the number of slices simultaneously excited, or $MB$ factor. Thus, SMS offers the possibility of increasing the spectral bandwidth of the fMRI time-series (\cite{risk2021multiband}) for fixed resolution levels whilst reducing the risk of signal contamination from physiological sources into the band of interest of resting state networks (RSNs) and/or of specific tasks. 

While in most cases a modest $MB$ factor (2 or 3) is sufficient to resolve respiratory-induced aliasing, cardiac aliasing is problematic even when a high $MB$ factor is used (\cite{aslan2019extraction}). For example, by taking as reference the fMRI data released as part of the the Human Connectome Project (HCP) (\cite{van2013wu}), which represents the state-of-the-art from an image acquisition standpoint, a $MB$ factor of 8 combined with $Z=72$ slices  resulted in a $TR$ of 0.72 seconds (\cite{uugurbil2013pushing}). Albeit fast, such acquisition is still vulnerable to aliasing of the first (and higher) cardiac harmonics in all subjects with a HR higher than 42 BPM. \\ This technological limitation may explain the rather large body of model-based work (\cite{chang2009influence, glover2000image, hu1995retrospective}) and the application of signal-processing techniques such as principal/independent component analysis (\cite{behzadi2007component, churchill2013phycaa+, perlbarg2007corsica, pruim2015ica, salimi2014automatic}) for the correction/understanding of such phenomenon.

Notably, there has also been work aiming at the correction of physiological noise when ultra-fast acquisitions ($TR<0.5$ seconds) are used (\cite{agrawal2020model, frank2001estimation, jahanian2019advantages}), although to achieve such short repetition times brain coverage and/or spatial resolution had to be compromised.

Most recently, a data-driven method to resolve the propagation of semi-periodic waves, which would otherwise alias into the fMRI time-series, has been proposed (\cite{voss2018hypersampling}). The hyper-sampling technique allows to resolve cerebral pulse waves of one heartbeat duration by exploiting \textit{i)} external physiological recordings of the cardiac signal and \textit{ii)} a retro-binning algorithm. While the cardiac signal is typically measured by means of photoplethysmogram (PPG) or electrocardiogram (ECG) devices, the reconstruction algorithm is designed to produce spatially resolved cardiac waveforms of one heartbeat duration. Building on this model, the Hypersampling by Analytic Phase Projection – Yay! (happy) technique (\cite{aslan2019extraction}) aims at extracting a single cardiac waveform for the entire duration of the fMRI exam. Happy does not rely on external recordings, but instead it exploits the multi-slice nature of the fMRI acquisition. Under the assumptions that the cardiac signal is ``\textit{i) pseudo-periodic, ii) somewhat coherent within any given slice, and iii) similarly shaped throughout the brain}'', authors were able to extract excellent estimates of the cardiac signal which closely resembled PPG recordings. Furthermore, it was possible to improve the shape of the cardiac waveform using a deep learning reconstruction filter (\cite{aslan2019extraction}).

In this study, we introduce the Data-driven WHOle-brain CArdiac signal REgression from highly accelerated simultaneous multi-Slice fMRI acquisitions (WHOCARES) method, demonstrating that it is feasible to temporally and spatially resolve cardiac waveforms throughout the whole brain at each voxel location. We propose to achieve this in a completely data-driven fashion, thus without external physiological recordings nor ad hoc modeling assumptions. In particular, when a SMS sequence is employed, the number of slices excited at different times reduces from $Z$ to $\frac{Z}{MB}$. Since in SMS the spatial distance of the slices simultaneously excited (i.e. those slices belonging to the same RF pulse) is maximized to minimize slice cross-talk and to reduce the burden on the CAIPIRINHA encoding gradients (\cite{setsompop2012blipped}), the set of $\frac{Z}{MB}$ temporal adjacent slices (i.e. those slices belonging to different RF pulses) are also confined within pre-defined regional slabs. Thus, WHOCARES capitalizes on the following observations: \textit{a)} that TS is sufficiently short to resolve the cardiac signal (\cite{aslan2019extraction}) and \textit{b)} when a ``sufficiently high'' $MB$ factor is employed, SMS confines slices acquired at different times to a compact region of space. In this study, we exploit these observations for the construction of a voxel-wise cardiac signal regressor. It is also shown that WHOCARES is capable of generating high-quality vessel maps. We developed the methodology on a sub-cohort of 774 healthy subjects selected from the HCP database. We also compare the proposed technique against the state-of-the-art (cardiac only) retrospective correction in the image domain (RETROICOR) method (\cite{glover2000image}). Note that we will refer to the cardiac part of RETROICOR as Cardio-RETROICOR. The source code of WHOCARES is freely available at \href{https://github.com/gferrazzi/WHOCARES}{https://github.com/gferrazzi/WHOCARES}.

\section{Methods} \label{section:Methods}

\subsection{Overview} \label{subsection:Methods:Overview}

\begin{figure}[t]
\centering
\includegraphics[scale=0.11]{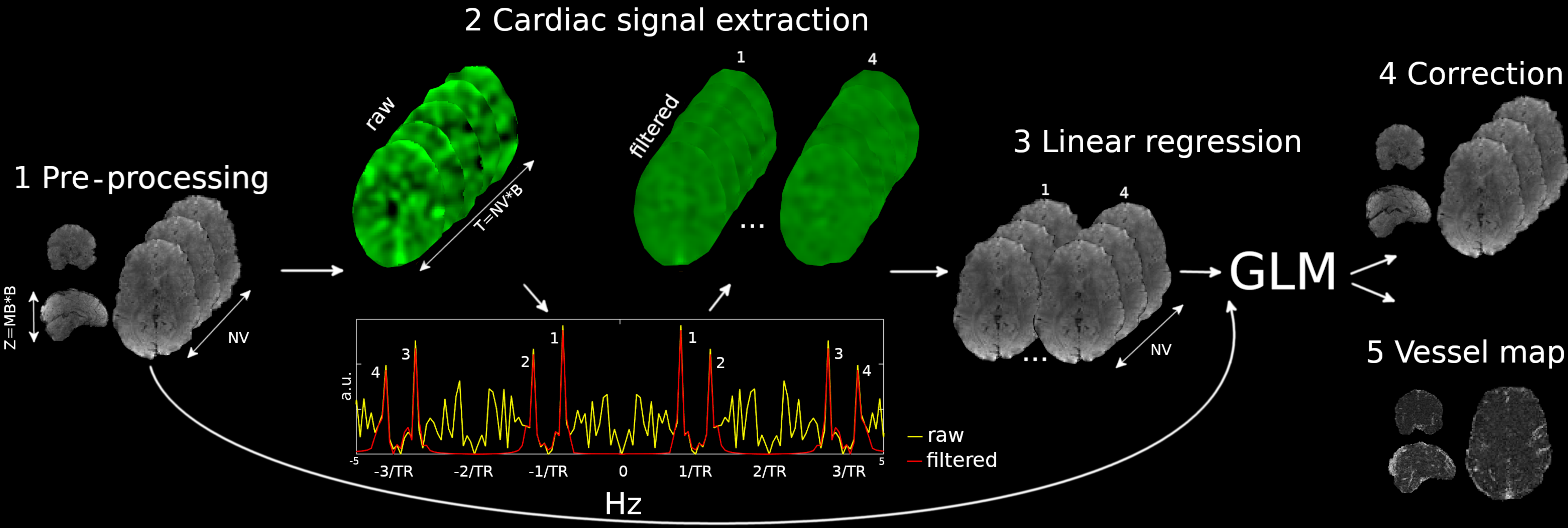}
\caption{\textit{WHOCARES pipeline}. Step 1: pre-processed data of size $Z$ and $NV$ along slice the temporal domains. Step 2, top left: reformatted hybrid dataset of size $MB \times T$ along third and fourth dimensions. Step 2, bottom: temporal Fourier transform of the hybrid time-series at a cardiac voxel (yellow) and the effect of filtering cardiac components 1 to 4 (red). Step 2, top right: filtered hybrid components 1 to 4. Step 3: GLM of the pre-processed data and the re-formatted cardiac signals. Steps 4 and 5: corrected fMRI dataset and vessel (MI) map between the pre-processed fMRI dataset and the cardiac regressor.}
\label{Figure1}
\end{figure}

We aim at the extraction of a cardiac signal regressor from the fMRI time-series. Figure~\ref{Figure1} outlines the main steps of WHOCARES. After data pre-processing (Figure \ref{Figure1}, step 1 - see section~\ref{prepro1} for details), the 4D fMRI dataset is processed to extract a cardiac signal regressor. If we label with $B=\frac{Z}{MB}$ the number of RF excitations per imaging volume (or equivalently, the number of slices acquired at different times within each of the $MB$ available blocks/slabs) and with $NV$ the number of temporal volumes, the pre-processed data is re-formatted along the third (slice) and fourth (temporal) dimensions resulting in a hybrid dataset of size $MB \times T$ with $T=NV \times B$ (step 2, top left). This effectively enhances the sampling rate of the fMRI scan from $TR$ to $TS=\frac{TR}{B}$. Note that some level of processing prior to re-formatting is required to remove the effect of the anatomy from the fMRI time-series. Please see section ~\ref{prepro2} for further details. A temporal Fourier transform of the hybrid dataset is computed at each voxel (step 2, bottom). This leads to signals of pseudo-periodic behaviour with main period $\frac{1}{TR}$ (step 2, bottom plot, yellow). Moreover, it comprises four super-resolved signal peaks (labelled as 1, 2, 3 and 4) that resemble cardiac pulsation (see section~\ref{prepro2} for a detailed explanation). These are isolated using a bank of bandpass filters centred on their corresponding frequencies (step 2, bottom plot, red). Extracted cardiac components 1, 2, 3 and 4 are re-formatted to match the spatial and temporal resolution of the original fMRI data (step 3, left). Note that the effect of the main anatomy is also re-introduced. A Generalized Linear Model (GLM) is used for the construction of a voxel-wise cardiac signal regressor by fitting components 1, 2, 3 and 4 onto the pre-processed fMRI dataset (step 3, right). Finally, cardiac pulsatility is regressed out from the fMRI time-series (step 4) and a vessel map is computed (step 5). The latter is calculated as the mutual information (MI) (\cite{duncan1970calculation}) between the pre-processed fMRI data and the computed regressor. Please see section~\ref{prepro3} for further details.

\subsection {Image Acquisition} \label{subsection:Methods:ImgAcq}
 
In the present work, we employed the un-preprocessed fMRI scans from the HCP S1200 release (\cite{van2013wu}). The scanning protocol, recruitment of participants and informed written consent were gathered by the University of Washington (Seattle, USA). Imaging was performed on a Siemens 3 Tesla (3T) Skyra MR scanner. Briefly, the data employed in this study consists of:

\begin{itemize}

    \item \textbf{fMRI data}: four resting state fMRI datasets were collected over two sessions (session number 1 and session number 2). Each session comprised two echo planar imaging (EPI) scans acquired using opposite phase encoding directions (Left-Right (LR) vs. Right-Left (RL)). In each run, $NV=1200$ time-frames were acquired using $MB=8$ at $2mm$ isotropic resolution and a $TR=0.72$ seconds. An odd/even slice order scheme was employed within each block/slab. Other parameters were: flip angle (FA) of  $52^o$ and an echo time $TE=33ms$. Further details can be found in (\cite{glasser2013minimal,uugurbil2013pushing}).
		
    \item \textbf{Physiological recordings}: PPG recordings at $400Hz$ including cardiac and respiratory traces were recorded as well.
    
\end{itemize}

\subsubsection {Subject selection} \label{subsection:Methods:SubSelect}

As noted recently (\cite{aslan2019extraction}), the failure rate of PPG recordings in the HCP is relatively high. Therefore, two sub-groups were selected from the HCP cohort. This was done according to recent work (\cite{orban2020time}), where authors provided quality control (QC) measurements of physiological recordings from the entire HCP cohort together with a system of queries to select sub-groups. Out of the 1200 subjects scanned twice during the two independent sessions, the following selection criteria were applicable for both groups: data had to belong to session number 1 only, the phase encoding direction had to be LR, PPG recordings had to be labelled as ``presen'', and the label ``high motion'' had to be classified as absent. The two subgroups differed only in relation to the result of the PPG cardiac signal QC test. In particular, the first and second group were formed by those subjects for which the QC test had a positive/negative outcome. We name these groups as ``good PPG'' and ``bad PPG'', respectively.

The application of the following criteria resulted in a cohort of 304 subjects for the good PPG group, and 476 subjects for the bad PPG group. However, because of corrupted PPG recordings, 1 subject from the good PPG group and 5 from the bad PPG group were excluded. Moreover, upon careful inspection of the results from the bad PPG group (see section \ref{section:Results} for details), it appeared clear that Cardio-RETROICOR had failed in a number of cases. These subjects were labelled and isolated forming the Cardio-RETROICOR failed group. Thus, final groups were as follows: 303 subjects formed the good PPG group, 392 the bad PPG group, and 79 the Cardio-RETROICOR failed group.

\subsection{Data Analysis} \label{dataanalysis}

The WHOCARES pipeline is formed by five independent steps  (see Figure \ref{Figure1}). Each of these will be discussed in details in subsections  \ref{prepro1}, \ref{prepro2} and \ref{prepro3}.
Section \ref{RETROcomp} describes the implementation of Cardio-RETROICOR, whereas Section \ref{Statistical} discusses how the two techniques were compared. Briefly, this was done by quantifying the MI shared between the computed regressors and the fMRI time series. 

\subsubsection{Pre-processing} \label{prepro1}

Standard pre-processing steps such as motion and/or distortion correction are likely to mix signals from different slices and so it is advisable to correct for these effects only after cardiac signal regression is performed (\cite{aslan2019extraction, voss2018hypersampling}). This is the reason why in this work the un-preprocessed raw data provided within the HCP distribution was employed. 
At first, a binary mask of the brain is retrieved using the BET tool of FSL (\cite{smith2002fast}) run on the fMRI temporal average. Subsequently, and as recently recommended (\cite{aslan2019extraction}), each voxel time-series is de-trended using a 3rd order polynomial function with Matlab (MathWorks, Natick, MA, United States). The application of these operations lead to the pre-processed dataset visible in Figure \ref{Figure1}, step 1.

\subsubsection{Cardiac signal extraction} \label{prepro2}

The first step comprises the application of a 2D spatial filter with a Gaussian kernel of Full Width Half Maximum (FWHM) of 1 pixel. The filter is applied on a slice-by-slice basis, as 3D filtering mixes signals from different slices. The effect of the anatomy is subsequently removed from the fMRI time-series. This is achieved by dividing the fMRI data by its temporal mean (note that outside the brain where no signal is present, a value of one is imposed). Each voxel time-course within the brain is subsequently normalized by its median absolute deviation (MAD) (\cite{aslan2019extraction}). This has the effect of equalizing the temporal variance of the time-series across slices, thus reducing the number of temporal spikes in the step described below.

The data is rearranged from a $Z \times NV$ matrix to a $MB \times T$ grid along third (slice) and fourth (time) dimensions. This effectively enhances the sampling rate of the fMRI scan from $TR$ to $TS=\frac{TR}{B}$. Note that, within each block/slab of $B$ slices, the data is read according to the slice acquisition order (odd/even), so the slices at each super-resolved timepoint in the block are at similar, but not identical locations. This operation is repeated within each block/slab separately, leading to the hybrid super-resolved dataset of Figure \ref{Figure1} (step 2, top left).

To reveal cardiac pulsatility, the temporal Fourier transform of the hybrid dataset is computed at each voxel location (Figure \ref{Figure1}, step 2, bottom) (\cite{biswal1996reduction}). However, since the acquisition of the HCP project lasts 14.4 minutes, fluctuations of the heartbeat are expected. Thus, to limit the extent to which HRV broadens cardiac peaks in the frequency spectrum whilst ensuring a sufficient spectral density of the Fourier transform, a finite number of $W=180$ non-overlapping frames (corresponding to temporal windows of 14.4 seconds) was empirically chosen from the super-resolved dataset. The following procedure is repeated until all $\frac{T}{W}$ segments are processed.

To suppress the DC component prior to Fourier transformation, the temporal mean is removed from each segment. Then, a Fourier transform is computed to extract the cardiac signal in the reshuffled space. When a group of voxels comprising vessels and/or arteries are sampled successively in time, a cardiac peak (peak number 1 in Figure \ref{Figure1}, step 2, bottom red) at approximately $1Hz$ emerges from the background noise (yellow). Note that the peak lies beyond the Nyquist limit set by the fMRI acquisition protocol, i.e. $\frac{1}{2 \times TR} \approx 0.69Hz$.

Although efforts were made to reduce spikes of temporal adjacent slices, each slice samples different parts of the brain and so it is inevitable that some level of spurious fluctuations is introduced. These fluctuations have a semi-periodic behavior since each anatomical location is sampled every $TR$ periods (Figure \ref{Figure1}, step 2, bottom plot, yellow)  (\cite{aslan2019extraction}). Thus, in addition to the super-resolved cardiac peak number 1, three cardiac replicas (peaks number 2, 3 and 4 of Figure \ref{Figure1}, bottom red) can be observed. These are placed symmetrically with respect to frequencies $\frac{1}{TR}$ and $\frac{3}{TR}$.

The following step aims at the extraction of each cardiac component. To compute the average HR within each segment, the happy toolbox (\cite{aslan2019extraction}) is employed (see footnote \footnote{The output of the happy toolbox consists of a single cardiac waveform re-sampled at $25Hz$ which resembles the PPG signal. The first step involves temporal filtering of the reconstructed cardiac waveforms between 25 and 150 BPMs. The cardiac signal within each segment of data is then extracted and the positions of the maximum values retrieved. To remove spurious maxima, the following constraints are applied after normalizing each cardiac waveform between -1 and 1; \textit{i})  only positive peaks (i.e. above the value 0.2) are counted; \textit{ii}) the minimum peak to peak distance has to be larger than 0.3 seconds and; \textit{iii}) first and last peaks within each segment of data are discarded. Peak to peak intervals are calculated as the relative distance (in seconds) between consecutive maxima. To improve robustness against spurious beats, values respectively below and above the 20\% and 80\% percentile of all intervals were removed. Finally, the median of all remaining intervals is calculated, and the HR (in $Hz$) per segment of data is computed as the inverse of such number. HR is calculated for each of the $\frac{T}{W}$ segments separately. To remove spurious estimations, a $5^{th}$ order polynomial function is fitted onto all HR values of all segments leading to a smooth curve.} for details). Once the average HR within each segment is available, the center frequency for peaks 1, 2, 3 and 4 is computed respectively as: $HR$, $\frac{2}{TR}-HR$, $\frac{2}{TR}+HR$ and $\frac{4}{TR}-HR$. A bank of passband filters of width $\pm 0.2Hz$ centered on each of these frequencies are employed to isolate cardiac components 1, 2, 3 and 4. Each cardiac component is subsequently Fourier transformed back to the temporal domain. After a mean value of one is imposed to each segment, these are compounded in time forming the set of four super-resolved cardiac regressors shown in Figure 1, step 2, top right.

\subsubsection{Linear regression, fMRI correction and calculation of vessel maps} \label{prepro3}

The four cardiac hyper-resolved components are re-formatted to match the original fMRI spatial and temporal resolutions. Note that the average anatomy is re-introduced at this stage (Figure \ref{Figure1}, step 3, left). This results in four cardiac signals defined at the $TR$ temporal resolution at each slice location. The final cardiac regressor is computed as the linear summation of the fit of these signals to the pre-processed data (Figure \ref{Figure1}, step 1) using a GLM (Figure \ref{Figure1}, step 3, right). 

Finally, cardiac pulsatility is regressed out from the fMRI time-series (step 4) and a vessel map (step 5) is created as the MI shared between the cardiac signal regressor and pre-processed fMRI data. MI is computed as the subtraction of two entropy terms. Specifically, we used the Gaussian Copula MI technique, which is a rank-based method resulting in a lower bound estimation of the ``true'' MI (\cite{ince2017statistical}). Note, finally, that the technique does not impose priors on the marginal distributions of the random variables.

\subsubsection{Comparison against Cardio-RETROICOR} \label{RETROcomp}

Cardio-RETROICOR (\cite{glover2000image, kassinopoulos2019identification}) was employed to retrieve cardiac pulsatility. To achieve this goal, the position of the cardiac peaks relative to the main fMRI acquisition is estimated from the PPG signal after bandpass filtering between 25 and 150 BPMs. A basis of three sine and cosine functions with main period set to be equal to, twice and three times shorter than the instantaneous beat to beat distance, respectively, is created and discretized at $TR$ intervals. The regressor is fitted independently at each voxel location (\cite{glover2000image}). In order to prevent bias towards any of the methods, the pre-processed data from Figure \ref{Figure1} (step 1) is used as input of Cardio-RETROICOR. Once the regressor is calculated, cardiac pulsatility is regressed out from the fMRI time-series and a vessel map (i.e. a MI map) is constructed for each subject independently as described in section \ref{prepro3}.

\subsubsection{Statistical comparisons} \label{Statistical}

Vessels masks were retrieved by considering the $95^{th}$ percentile value from the voxel intensity distributions of the MI maps. This was done both for Cardio-RETROICOR and WHOCARES. The resulting vessel masks were then multiplied together to highlight spatial locations labelled as ``cardiac'' locations by both procedures. The performance of the two techniques was then compared by selecting the average MI score within these regions. We also investigated whether a relationship between the average MI score and average BPM across the entire fMRI exam exists. The same procedure was repeated for HRV. For a detailed description on how BPM and HRV were calculated, please refer to footnote \footnote{The average BPM was calculated from the output of happy similarly to HR, although a single value per fMRI exam (as opposed to per segment $W$) was obtained. After calculating the instantaneous beat to beat distance using the same constrains as for HR (see previous footnote), the median of the resulting scores was taken, and the average BPM was computed as the inverse of such number multiplied by 60. HRV was calculated as the standard deviation (std) of the temporal derivative of the instantaneous beat to beat distance. Note therefore that such quantity is expressed in seconds}. To compare dependencies of these scores, the angle from the linear fit of the scatter plot BPM/HRV vs. average MI was computed across methods/groups, after normalization of BPM and HRV values to the dynamic range set by the average MI scores. We name such scaled quantities as ``normalized BPM'' and ``normalized HRV''. 

For the statistical analysis, comparisons of the average MI scores between methods and across groups were performed. Lilliefors tests were used to assess whether the MI scores within the vessel masks were normally distributed. When this criterion was fulfilled ($p\geq 0.001$), paired/unpaired t-tests were run between distributions across methods/groups, respectively. If the criterion of non Gaussianity was not fulfilled ($p < 0.001$), non parametric Wilcoxon signed-rank/rank tests were run between methods/groups, respectively. To test whether the std of the distributions were the same across methods/groups, the Levene test was employed.  In all tests, p-values less than $0.001$ indicated statistical significance.

\section{Results} \label{section:Results}

\begin{figure}[t]
\centering
\includegraphics[scale=0.065]{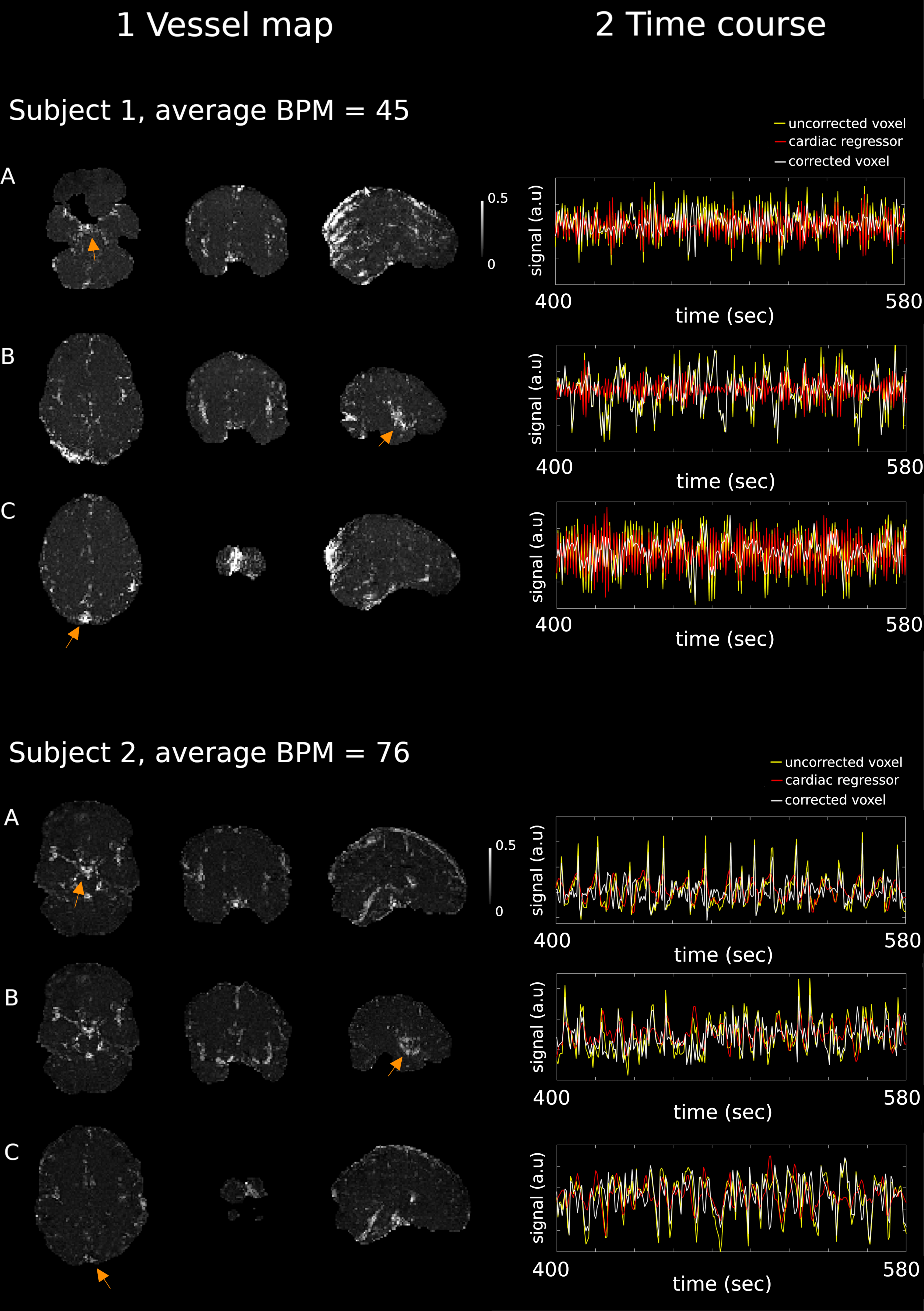}
\caption{\textit{Results on two representative subjects}. Panel 1, left: MI/vessel maps between the pre-processed fMRI data and the cardiac regressor for WHOCARES in two exemplary subjects (top vs bottom plots). (A) circle of Willis, (B) middle cerebral artery and (C) sagittal sinus regions. Panel 2, right: Three-minute extract of the fMRI pre-processed time-series (yellow), the calculated cardiac signal regressor (red) and the corrected fMRI data (white). The spatial location of the selected voxels is indicated by the orange arrows of Panel 1.}
\label{Figure2}
\end{figure}

Figure \ref{Figure2} (left) shows MI maps between the pre-processed fMRI time-series and the computed cardiac signal regressors in two representative subjects (top vs bottom plots). In both cases, MI is maximum in proximity to blood vessels and arteries. This includes, although it is not restricted to, the (A) circle of Willis, (B) the middle cerebral artery and (C) the sagittal sinus. Figure \ref{Figure2} (right) shows three-minute extracts from the fMRI pre-processed time-series (yellow), the calculated cardiac signal regressors (red) and the corrected fMRI data (white). The voxel time-courses from which the extracts were taken are highlighted on the left panels with orange arrows. While cardiac activity induces high-frequency noise in subject number 1, slower fluctuations are observed in subject number 2. This is consistent when looking at the average HR in these subjects; in the first case, this was 45 BPM and therefore the perceived normalized aliased frequency is 0.46 (relative to a Nyquist frequency of 0.5). In the second case, average BPM was 75 and the perceived normalized frequency 0.1. The application of WHOCARES reduces high/low-frequency noise caused by cardiac aliasing, although residual fluctuations are present. In general, MI for subject 1 was higher than subject 2.

\begin{figure}[t]
\centering
\includegraphics[scale=0.3]{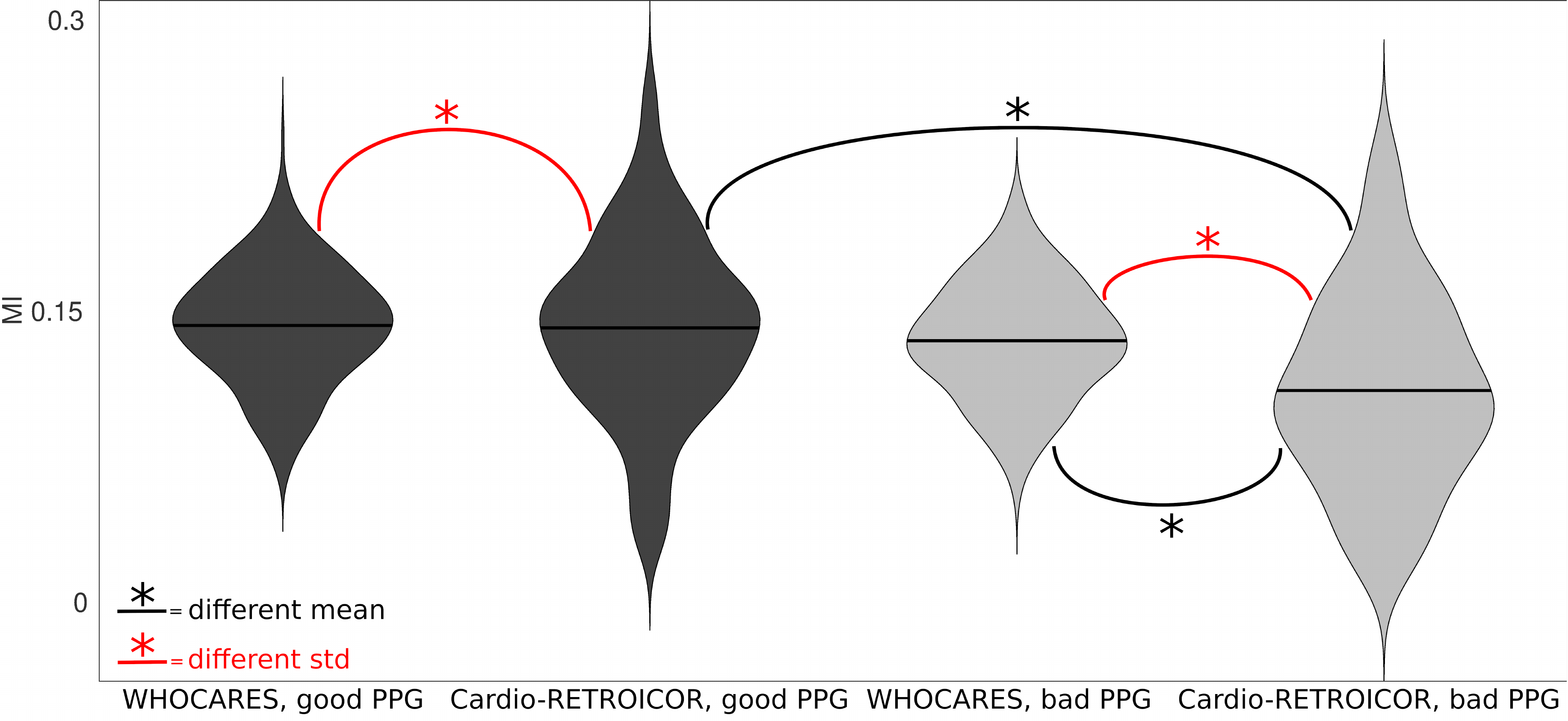}
\caption{\textit{Violin plots of WHOCARES vs. Cardio-RETROICOR performances}. Violin plots of the average MI distributions of good/bad PPG groups (dark/light grey) for WHOCARES (left) and Cardio-RETROICOR (right). Note that in this representation subjects belonging to the Cardio-RETROICOR failed group were removed.}
\label{Figure3}
\end{figure}

Figure \ref{Figure3} reports violin plots of the average MI distributions of good/bad PPG groups (dark/light grey) for WHOCARES (left) and Cardio-RETROICOR (right). Note that in this representation subjects belonging to the Cardio-RETROICOR failed group were removed. For WHOCARES, average MI distributions across groups were remarkably similar, although the p-value highlighting differences of the mean was almost significant ($0.143$ vs. $0.136$, $p=0.005$). Note that the std of the two distributions was, instead,  the same ($0.03$ vs. $0.03$, $p=0.706$). The performances of Cardio-RETROICOR worsened considerably when comparing good PPG group vs. bad PPG group mean values ($0.141$ vs. $0.108$, $p \approx 0$). However, stds were similar ($0.047$ vs. $0.05$, $p=0.053$). In the good PPG group, WHOCARES and Cardio-RETROICOR resulted in the same average MI ($0.143$ vs. $0.141$, $p = 0.343$), although the distribution of Cardio-RETROICOR was more widespread in std ($0.03$ vs. $0.047$, $p \approx 0$). Finally, in the bad PPG group, WHOCARES reached better performances than Cardio-RETROICOR both in terms of mean ($0.136$ vs. $0.108$, $p \approx 0$) and std ($0.03$ vs. $0.05$, $p \approx 0$).

\begin{figure}[t]
\centering
\includegraphics[scale=0.5]{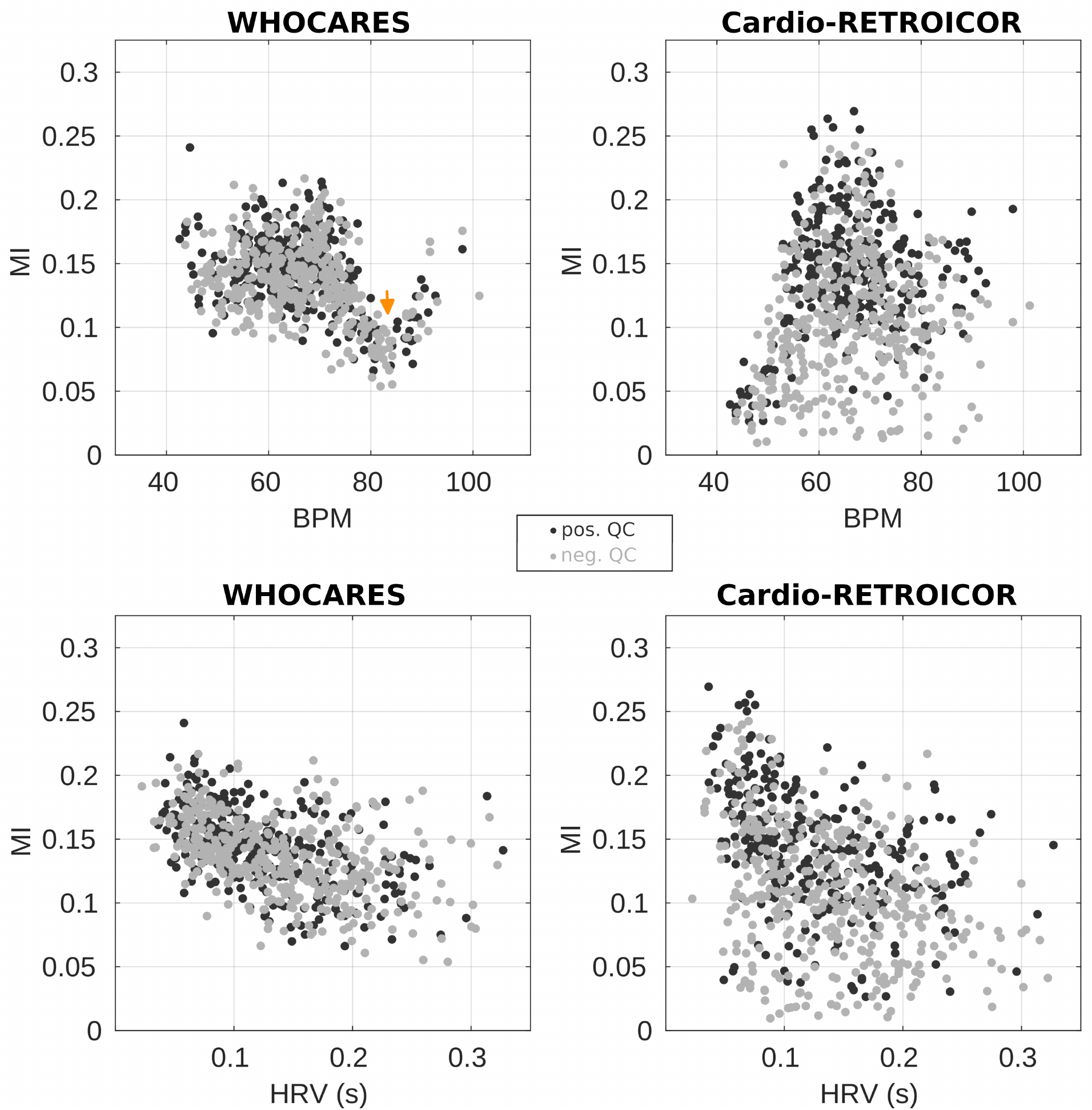}
\caption{\textit{Performances of WHOCARES vs. Cardio-RETROICOR}. Relationship between the average MI score against BPM (top) and HRV (bottom) for the good PPG group (dark grey) and the bad PPG group (light grey) of WHOCARES (left) vs. Cardio-RETROICOR (right). Note that in this representation subjects belonging to the Cardio-RETROICOR failed group were removed.}
\label{Figure4}
\end{figure}

Figure \ref{Figure4} reports the relationship between average MI and BPM / HRV (top / bottom) for good PPG group (dark grey) and bad PPG group (light grey) of WHOCARES (left) vs. Cardio-RETROICOR (right). As in Figure \ref{Figure3}, subjects belonging to the Cardio-RETROICOR failed group were removed. In the good/bad PPG groups, there was a non-linear trend of the average MI against BPM for WHOCARES. This is visible in Figure \ref{Figure4}, where a dip in average MI values can be observed between 75 and 90 BPM (arrow). Note that the result is consistent and reproducible across groups. The reason/remedy of/for such behaviour is described in details in section \ref{subsection:Discussion:limitations}. However, note that when this source of non-linearity is removed (by removing data-points between 75 and 90 BPM), the relationship between variables becomes linear, and the performances of WHOCARES relatively independent from the normalized BPM (angle formed by the linear fit of average MI vs. normalized BPM of $-4^o$ and $5^o$in good/bad PPG groups, respectively). For Cardio-RETROICOR, there was a positive trend (angle from linear fit of $17^o$ and $16^o$ degrees for good/bad PPG groups, respectively) between average MI and the normalized BPM. In relation to the normalized HRV, there was a negative relationship between the average MI both in WHOCARES ($-19^o$ and $-21^o$ degrees for good/bad PPG groups, respectively) and Cardio-RETROICOR ($-22^o$ and $-24^o$ degrees for good/bad PPG groups, respectively).

\begin{figure}[t]
\centering
\includegraphics[scale=0.12]{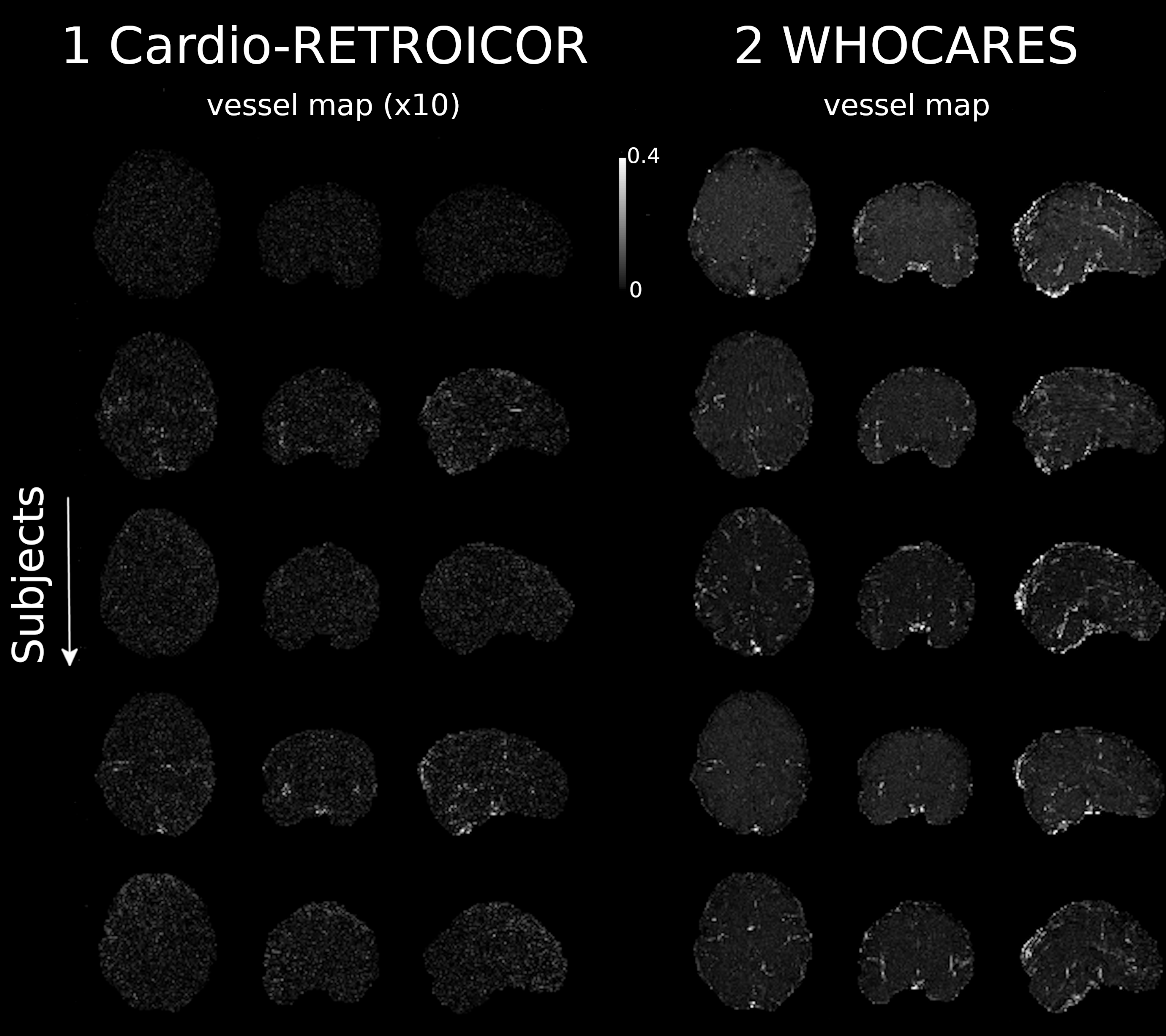}
\caption{\textit{Qualitative comparison of WHOCARES vs. Cardio-RETROICOR in Cardio-RETROICOR failed group}. Vessel maps from 5 subjects chosen at random reconstructed using Cardio-RETROICOR (left, MI maps multiplied by 10) and WHOCARES (right).}
\label{Figure5}
\end{figure}

79 subjects (Cardio-RETROICOR failed group) were found not to exhibit structure within the MI maps corresponding to vessels/brain arteries using Cardio-RETROICOR (failure rate of 10\%).  Figure \ref{Figure5} shows 5 subjects chosen at random from this group. No apparent MI within vessels and/or arteries can be observed using Cardio-RETROICOR (note that in Figure \ref{Figure5}, MI maps obtained with Cardio-RETROICOR were multiplied by 10 to improve readability). In contrast, WHOCARES is generally capable of retrieving vessel maps (right).

\section{Discussion}  \label{subsection:Discussion}

Cardiac pulsation is a physiological confound of fMRI analysis pipelines that introduces spurious fluctuations of the BOLD signal. To overcome cardiac aliasing associated with a limited temporal resolution in fMRI, we developed a data-driven technique to temporally and spatially resolve cardiac signals from the BOLD signal itself, i.e. without the need of acquiring external physiological recordings. We sought to achieve this using a data-driven strategy, thus without imposing modeling priors on the shape of the regressor. This is achievable by recognizing that the time between consecutive excitations, rather than the time between the acquisition of consecutive volumes, is the natural clock of the system (\cite{aslan2019extraction, voss2018hypersampling}), and by combining such principle with highly accelerated SMS data (\cite{glasser2013minimal, larkman2001use, uugurbil2013pushing}). By inferring cardiac signal contributions from the fMRI data itself, cardiac noise was found to be spatially localized, especially in and around blood vessels (Figure \ref{Figure2}), in line with previous literature based on mathematical modeling of individual cardiac responses (\cite{glover2000image}). 

WHOCARES is compatible with existing fMRI processing pipelines. In particular, the computed regressor can be paired with other regressors aiming at modeling physiological noise of different nature, signal drifts, motion, task design etc. Note that, in this context, since the GLM is typically performed only after a number of image-registration processing steps (i.e. distortion correction, motion correction, normalization to a standard template etc.) are applied to the fMRI data, there is the need to project the regressor to the same working space. This can be done by applying the same set of transformations computed for the fMRI data to components 1, 2, 3 and 4 of Figure \ref{Figure1}, step 3. 

\subsection{Comparison between WHOCARES and Cardio-RETROICOR} \label{subsection:Discussion:cardio-RETROICOR}

Most fMRI studies focusing on cardiac pulsatility have employed external physiological recordings to model the effect of cardiac noise (\cite{glover2000image, kasper2017physio, kassinopoulos2021physiological, kelley2008automatic}). As such, these methods rely on the quality/presence of a cardiac trace. This is visible in Figure \ref{Figure3}, where a dependency between the performances of Cardio-RETROICOR and the quality of PPG recordings exist. Instead, WHOCARES was relatively stable across groups, and so whilst the performances against Cardio-RETROICOR were comparable for the good PPG group, the amount of information shared between the regressor and the fMRI time-series was superior in the bad PPG group. Moreover, visual inspection of the vessel maps obtained with Cardio-RETROICOR revealed a sub-group (Cardio-RETROICOR failed group) in which the algorithm failed remarkably (Figure \ref{Figure5}, left). Failure rate for Cardio-RETROCOIR was relatively high (approximately 10\%). Tracing back the reason of such failure revealed that the cardiac peak detection step from the PPG trace had failed, since the quality of the latter was severely compromised. On the other hand, WHOCARES allowed for the retrospective correction of such data (Figure \ref{Figure5}, right). This has numerous applications; for example at ultra-high-field ($ \geq 7T$), where the effect of cardiac noise is more prominent and the ability to measure physiological traces is often hindered (\cite{stab2016ecg}).

Another aspect related to external physiological recordings is linked to the observation that cardiac systole originating from the heart reaches the fingers only after a certain delay (\cite{allen2007photoplethysmography}). Note that PPG signals themselves are shifted relative to systolic pulses entering the brain. Therefore, when extracting the positions of the cardiac peaks from PPG recordings, it is inevitable to introduce temporal shifts between Cardio-RETROICOR and the fMRI data. While part of the originating uncertainties is thought to be resolved by the fact that the sine and cosine components of Cardio-RETROICOR take care of all but the most severe time-shifts, it has been suggested that incorporating such shifts into a mathematical model can actually yield to improved results (\cite{kassinopoulos2021physiological}). In this context, note that the cardiac waveforms retrieved with happy are also shifted relative to the PPG signal (\cite{aslan2019extraction}). However, these are employed for the sole purpose of retrieving the average HR within each segment of data, and so by construction there is no temporal discrepancy between the proposed regressor and the fMRI time-series.

Concerning the modeling, an open issue with Cardio-RETROICOR is linked to the choice of the optimal order of the model basis set, which may result in the over-fitting of the fMRI signal, thus carrying the risk of removing signal of interest (\cite{harvey2008brainstem}). Although these aspects have been partly addressed with a refined model that took into account the time-lag and the model order (\cite{kassinopoulos2021physiological}), model-based methods are not suited to describe physiological processes outside their mathematical domain. For example, initial evidence suggests that the explained variance of cardiac- and respiratory-induced fluctuations is less spatially localized than in RETROICOR (\cite{caballero2017methods}). In the past, research has also been conducted to characterize regional respiratory and cardiac response functions (\cite{birn2006separating, birn2008respiration, chang2009influence, chen2020resting, kassinopoulos2021physiological}) and, in this context, the data-driven nature of WHOCARES may be of interest, as it does not impose modeling priors.

\subsection{Limitations and future work}  \label{subsection:Discussion:limitations}

Our study has several limitations. By reformatting the data along the through-slice direction, WHOCARES is predominantly sensitive to ``vertical flux'' of the blood, i.e. the fresh blood coming from the carotid arteries into the brain and back through the the jugular veins, rather than ``in-plane flux'' components, i.e. from the left to the right and vice-versa. 

Although WHOCARES is capable of recovering the first harmonic of the cardiac signal - whose power is split into four sub-peaks (see Figure \ref{Figure1}, step 2, bottom red) - we did not find evidence of higher harmonics in the recovered temporal Fourier transforms. This aspect - which differs substantially from the approach adopted in happy, where spatial averaging within vessels and brain arteries allows for a more accurate spectral representation of the cardiac signal including higher order harmonics (\cite{aslan2019extraction}) - is challenging, since the act of averaging \textit{i}) prevents the capability of resolving the cardiac signal locally and \textit{ii}) mixes signals from distant brain regions so that differences in the blood arrival time cannot be taken into account.

Since the cardiac regressor is built from the fMRI data itself, a spatial Gaussian filter is applied to reduce spurious fluctuations outside neuro-vascular coupled regions. However, this also reduces MI values within brain arteries (data not shown). Thus, the performance of WHOCARES would benefit from an adaptive filtering technique where spatial smoothing is applied prominently outside vessels. In this context, a possible strategy would be to employ vessel maps retrievable through hyper-sampling (\cite{aslan2019extraction, voss2018hypersampling}) to guide spatial smoothing. 

Temporal filtering is applied in the reformatted space using a bandpass filter centred on the average HR with width $\pm 0.2Hz$ ($= \pm 12$BPM) (Figure \ref{Figure1}, step 2, bottom). With such procedure, we assume that \textit{i}) the estimated HR is an accurate representation of the true heart-beat frequency and \textit{ii}) the HR fluctuations within each segment of data are always within the spectral width of the filter. To meet these requirements, we extracted segments of 14.4 seconds, as opposed to the entire scan, and processed them sequentially to compute the cardiac regressor. However, other choices are possible, which may explain why there was a negative relationship between MI scores and HRV using our method (Figure \ref{Figure4}, bottom left).

Other limitations are linked to the technological requirements of fMRI acquisitions, including the use of highly accelerated SMS scans. In fact, a prerequisite of our technique is that the signal from temporally adjacent slices is well confined within predefined regions of space (or slabs). As the number of slices excited at different times is inversely proportional to the $MB$ factor, we expect WHOCARES to benefit greatly from higher acceleration. However, there is a limit to the highest possible multiband factor which is set by the encoding capabilities of the multiple receiver coils. 

In the HCP, an odd/even slice acquisition order was selected, possibly aiming at the minimization of slice cross-talk and/or to limit the extent of spin history artefacts if motion occurred (\cite{ferrazzi2014resting}). However, the spatio-temporal continuity of the slices is a desired property of WHOCARES which is expected to benefit from an ascending/descending (rather than an odd/even) slice acquisition scheme.   

Another relevant aspect emerges when looking at the dependency between the MI scores and BPM values using the proposed technique. In particular, a dip in the MI can be observed between 75 and 90 BPM, both in  the good/bad PPG groups (Figure \ref{Figure4}, arrow). When the average HR approaches this range, super-resolved cardiac peaks 1-2 and 3-4 of Figure \ref{Figure1} get closer in groups of two. Actually, with an instantaneous BPM of exactly $\frac{60}{TR} \approx \frac{75+90}{2} \approx 83.3$, peaks 1-2 and 3-4 overlay entirely (note that the band 75-90 BPM corresponds roughly to the width of the temporal filter – i.e. $\pm12BPM$). Thus, when trying to extract cardiac pulsatility at these frequencies, the information enclosed within pairs of regressors is similar, and so overall performances tend to drop, for then increasing again at higher BPMs. Nonetheless, WHOCARES showed a stable performance independent from the average heart-rate when subjects with a BPM between 75 and 90 were removed from the analysis (see section \ref{section:Results}). For all these reasons, and since none of the 774 subjects considered in this study exhibited heart-rates beyond 100 BPM, we predict the performances of WHOCARES to stabilize over the physiological range of the normal heart-rate at rest if a slightly faster acquisitions is employed (i.e. $TR<0.6$ seconds). However, such adjustment may require higher multiband factors and/or a slight drop in resolution. In contrast, the performances of Cardio-RETROICOR are positively correlated with BPM irrespective of $TR$ (Figure \ref{Figure4}), and understanding the reason of such behaviour (and possibly its resolution) is not straightforward.   

Regarding future work: research should be oriented towards the extension of WHOCARES to task-based fMRI, where physiological processes might be modulated by the presence of the experimental design  (\cite{glasser2018using}) and vice-versa. In this context, our approach could help disentangling physiologically-driven from neuronally-linked fluctuations, which is perhaps useful in the definition of physiological networks (\cite{chen2020resting}, \cite{xifra2021physiological}). From a clinical perspective, cardiac pulsatility has recently shown its application in the definition of pathophysiological biomarkers and monitoring of disease progression in age-related neurodegenerative disorders (\cite{kim2021cardiac}), which might benefit from WHOCARES implementation. Another example may be the case of stroke patients where, sometimes, the heart-rate shows pathological-related modifications and changes in cardiac pulsatility (\cite{geurts2019higher}). In this context, the inclusion of WHOCARES may help disentangling neuronally-linked from cardiac-related signal changes in fMRI studies.

\section{Conclusions}  \label{subsection:Conclusions}

Hyper-sampling and SMS imaging can be employed to spatio-temporally resolve cardiac waveforms in fMRI. WHOCARES holds basis for reliable mapping of the cardiac activity in the brain and for the construction of high-quality vessel maps. WHOCARES does not make specific assumptions on the shape of cardiac pulsation, it is independent from external physiological measurements, and it can be used for the retrospective correction of fMRI recordings when these are not available. The approach has been validated against the state-of-the-art RETROICOR method, achieving similar performances in terms of overall shared information with the fMRI time-series. WHOCARES has also proven stable over a wide range of heart-rates and heart-rate variability with respect to RETROICOR, especially when the quality of physiological recordings was compromised.

\section*{Acknowledgements}

This work was supported by the Italian Ministry of Health GR-2019-12368960 to GP, by the Research Foundation Flanders (FWO) (postdoctoral fellowship 1211820N to MM), and by the US National Institutes of Health grant numbers R01 NS097512 and R21 AG070383 (BBF). DM was funded by the Faculty of Psychology and Educational Sciences of Ghent University, and by the Research Foundation Flanders (FWO) (sabbatical bench fee K802720N).

	Data used in the preparation of this work were obtained from the MGH-USC Human Connectome Project (HCP) database (\href{https://ida.loni.usc.edu/login.jsp}{https://ida.loni.usc.edu/login.jsp}). The HCP project (Principal Investigators : Bruce Rosen, M.D., Ph.D., Martinos Center at Massachusetts General Hospital; Arthur W. Toga, Ph.D., University of Southern California, Van J. Weeden, MD, Martinos Center at Massachusetts General Hospital) is supported by the National Institute of Dental and Craniofacial Research (NIDCR), the National Institute of Mental Health (NIMH) and the National Institute of Neurological Disorders and Stroke (NINDS). Collectively, the HCP is the result of efforts of co-investigators from the University of Southern California, Martinos Center for Biomedical Imaging at Massachusetts General Hospital (MGH), Washington University, and the University of Minnesota. HCP data are disseminated by the Laboratory of Neuro Imaging at the University of Southern California.

\section*{Data and code availability statement}

\begin{itemize}
  \item The source code of WHOCARES together with the documentation specifying software requirements etc. is available at \\~\href{https://github.com/gferrazzi/WHOCARES}{https://github.com/gferrazzi/WHOCARES}.
  \item The documentation of the happy toolbox can be found at \\~\href{https://rapidtide.readthedocs.io/en/latest/index.html}{https://rapidtide.readthedocs.io/en/latest/index.html}. The source code at \href{https://github.com/bbfrederick/rapidtide}{https://github.com/bbfrederick/rapidtide}.
  \item The toolbox for computing Gaussian Copula Mutual Information can be found at~\href{https://github.com/robince/gcmi}{https://github.com/robince/gcmi}.
  \item FSL can be found at~\href{https://fsl.fmrib.ox.ac.uk/fsl/fslwiki/FslInstallation}{https://fsl.fmrib.ox.ac.uk/fsl/fslwiki/FslInstallation}.
	
\end{itemize}

\section*{Competing Interests Statement} 

All authors have no competing interests to declare.

\bibliography{mybibfile}

\end{document}